\documentclass{mn2e}
\usepackage{psfig}
\usepackage{mnras_cite}
\begin{document}

\title[Non-Gaussianities \& Early Reionization]
{Can Non-Gaussian Cosmological Models Explain the
WMAP's High Optical Depth for Reionization?}
\author[Chen et al]{Xuelei Chen$^1$, Asantha Cooray$^2$, Naoki Yoshida$^3$, Naoshi Sugiyama$^3$\\
$^1$Institute for Theoretical Physics, Kohn Hall, University of California, Santa Barbara, CA 93106\\
$^2$Theoretical Astrophysics, Mail Code 130-33, Caltech, Pasadena, CA
  91125, USA\\
$^3$Division of Theoretical Astrophysics, National Astronomical Observatory Japan, Mitaka, Tokyo 181-8588, Japan}

\maketitle

\begin{abstract}

The first-year Wilkinson Microwave Anisotropy Probe data 
suggest a high optical depth for Thomson scattering of 0.17 $\pm$ 0.04,
implying that the universe was reionized at an early epoch, $z\sim 20$. 
Such early reionization is likely to be caused by UV photons from 
first stars, but it appears that the observed high optical depth
can be reconciled within the standard structure formation 
model only if star-formation in the early universe was 
extremely efficient.  With normal star-formation efficiencies, 
cosmological models with non-Gaussian density fluctuations
may circumvent this conflict as high density peaks collapse at an earlier epoch than in
models with Gaussian fluctuations. We study cosmic reionization in non-Gaussian models 
and explore to what extent, within available constraints,
non-Gaussianities affect the reionization history. 
For mild non-Gaussian fluctuations at redshifts of 30 to 50,
the increase in optical depth 
remains at a level of a few percent and  appears unlikely to aid 
significantly in explaining the measured high optical depth.
On the other hand, within available observational constraints,
increasing the non-Gaussian nature of density fluctuations 
can easily reproduce the optical depth and may remain viable in underlying
models of non-Gaussianity with a scale-dependence.
\end{abstract}

\section{Introduction}

The excess large scale polarization signal in the 
Wilkinson Microwave Anisotropy Probe (WMAP) data has allowed the first
measurement of the optical depth to re-scattering by electrons with 
a value of $\tau = 0.17 \pm 0.04$ (Kogut et al. 2003). 
If interpreted as a sharp transition to a reionized universe from a 
neutral one, the redshift at which this transition happens is $z_{r}=17 \pm 5$
\cite{kogut,spergel}. Theoretical studies based on Cold
Dark Matter (CDM) models suggest that such a high reionization redshift 
--- beyond what was previously suggested with Gunn-Peterson troughs in 
$z \sim 6$ quasars from the Sloan survey (e.g., Fan et al. 2002) --- 
generally requires significant star-formation early on, at $z>10$
(Cen 2003; Haiman \& Holder 2003; Sokasian et al. 2003; Yoshida et al. 2003).

While conventional models of reionization based on ordinary stellar populations
(Population II stars) formed in galaxies alone are still consistent with the 
measured optical depth (Chiu, Fan \& Ostriker 2003; Somerville \& Livio 2003),
these models are not capable of explaining the high end of the allowed range 
for the optical depth.
The best fit value of $\tau=0.17$ is only achieved in models where
reionization takes place with the maximal efficiency in terms of 
star-formation and associated ionizing photon production 
(Fukugita \& Kawasaki 2003; Haiman \& Holder 2003; Cen 2003; 
Sokasian et al. 2003; Wyithe \& Loeb 2003). 
Alternatively, with a combination of a high photon escape fraction 
and a top-heavy IMF in early galaxies, the resulting total optical depth
can be increased to be marginally consistent with the WMAP data 
(e.g., Ciardi, Ferrara \& White 2003).
Although recent theoretical studies consistently indicate that the first 
stars formed out of a chemically pristine gas were rather massive 
(Abel, Bryan \& Norman 2002; Bromm, Coppi \& Larson 2002), 
hence supporting at least the latter condition, 
it is unclear whether such conditions are realized in primeval galaxies 
over a sufficiently long period. 
It is therefore intriguing to consider alternative mechanisms 
that do not require considerably high efficiency in star-formation or in 
ionizing photon production rate. 
Interestingly, such a scenario is possible even 
within the standard cosmological models.
For example, as we show later in the present paper, with $\sigma_8$ equal to 
or larger than unity, the optical depth to reionization can reach within the
range of the WMAP result
since with a higher normalization, halos collapse slightly 
at earlier times and are more numerous at a given epoch when compared to 
models with a lower value for the normalization of the matter power spectrum.

In light of this, with currently favored values for cosmological parameters,
invoking non-Gaussianities in primordial density fluctuations may be an 
interesting possibility and deserves further study. In the non-Gaussian 
models, one finds an increased fraction of high density peaks relative to Gaussian models, 
and these peaks are expected to collapse earlier than in the standard case
(Peebles 1997; White 1998; Koyama, Soda \& Taruya 1999; Robinson \& Baker 2000;
Willick 2000; Seto 2001; Mathis 2002). 
If star-formation is triggered in these collapsed halos, photons emitted from them 
may be able to, at least partially, reionize the universe very early on, 
causing a large total Thomson optical depth.

Excess non-Gaussian fluctuations generally lead to formation of structures 
at late times and could overproduce the cluster abundance at low redshifts 
(Willick 2000). However, the constraint obtained from cluster abundance
is rather uncertain, and previous studies based on models of primordial 
non-Gaussianities (e.g., Verde et al. 2001) suggest that, in certain models 
where non-Gaussianities are associated with the density field, even a small
primordial non-Gaussianity can significantly affect the number counts of objects 
at high redshifts. Avelino \& Liddle (2003) recently studied the evolution
of the mass fraction in collapsed halos in non-Gaussian models 
and calculated a simple lower estimate for the redshift of reionization.

In this {\it Letter}, we use a detailed model for the mass function 
in non-Gaussian models and compute the reionization history for some cases.
To this end we compute the total Thomson optical depth as a function of redshift.
We conclude that such non-Gaussian models, consistent with current constraints, 
can easily explain WMAP's results. 
We also argue that, while these models may eventually be constrained with large scale 
structure observations, there is still some freedom to obtain the necessary non-Gaussianity 
with models which are strongly scale-dependent. 

\begin{figure}
\centerline{\psfig{file=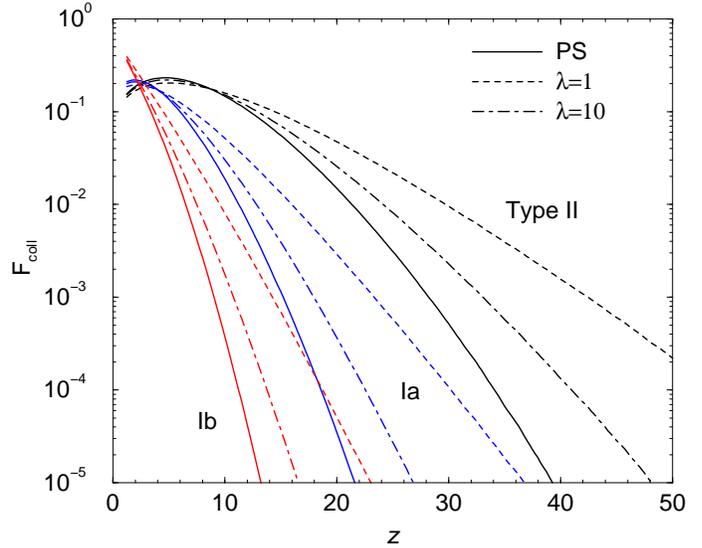,width=90mm,angle=-90}}
\caption{The fraction of mass in collapsed halos, between various mass limits (see Section~2 for a summary), 
as a function of redshift for the standard Press-Schechter mass function and for the non-Gaussian mass function
that includes a free parameter $\lambda$ that accounts for the departure from Gaussianity. Small values for
$\lambda$ represent a significant non-Gaussianity while $\lambda > 100$ is consistent with Gaussian 
fluctuations in the density field.
\label{fig:Fcoll}}
\end{figure}

\section{Calculational Method}

Following Haiman \& Holder (2003) (hereafter HH), we study star formation and
reionization using a simple toy model. 
Throughout the paper we assume the best fit flat $\Lambda$CDM model with 
power law fluctuations for the WMAP data:
$(\Omega_{m}, \Omega_{\Lambda}, \Omega_{b}, h, \sigma_{8}, n_s)=(0.27,
0.73, 0.044, 0.72, 0.9, 0.99)$.
Given these cosmological
parameters, we calculate the collapsed fraction of baryons using
a spherical halo model. HH distinguished
three different types of halos based essentially on
the halo mass, which captures
different nature of stars formed in them.
The collapsed halo fractions for each halo type is
\begin{equation}
F_{\rm coll} = \frac{1}{\bar{\rho}}
\int_{m_l}^{m_h} dm\; m n(m,z) \,,
\end{equation}
where $m_l$ and $m_h$ take values based on the gas temperature within
the halo such that
\begin{eqnarray}
m(z) &=& 0.94 \times 10^8 M_{\sun} \left(\frac{h}{0.7}\right)^{-1}
\left(\frac{\Omega_m}{0.3}\right)^{-1/2} \nonumber \\ &&\quad \times
\left(\frac{1+z}{10}\right)^{-3/2} \left(\frac{\mu}{0.6}\right)^{-3/2}
\left(\frac{T}{10^4 K}\right)^{3/2}\, ,
\end{eqnarray}
where $\mu$ is the mean molecular weight: in the case of ionized gas
with $T \geq 10^4$ K, $\mu=0.6$ and for neutral gas with $T < 10^4$ K,
$\mu=1.2$. The collapse fractions are divided to three halos between
100 K and 10$^4$ K (called Type II where molecular Hydrogen cooling
dominates), between 10$^4$ K and $2 \times 10^5$ K (Type Ia), and
above $2 \times 10^5$ K (Type Ib).  In both Type I halos, the dominant
cooling mechanism to form stars involves atomic hydrogen lines.

Using the collapsed mass fractions, we can now calculate the ionized HII
filling factor, which is the same as the mean electron fraction
$\bar{x}_e(z)$, by integrating for ionization produced by collapse
at all higher redshifts,
\begin{eqnarray}
\nonumber
F_{\rm HII}(z)&=&\rho_{\rm b}(z)
\int_{\infty}^{z}dz^{\prime}
\left\{
  \epsilon_{\rm Ib} \frac{dF_{\rm coll}^{\rm Ib}}{dz} (z^{\prime}) +
  \left[1-F_{\rm HII}(z^\prime)\right]\right.\times\\ &&\left.\left[
  \epsilon_{\rm Ia} \frac{dF_{\rm coll}^{\rm Ia}}{dz} (z^{\prime}) +
  \epsilon_{\rm II} \frac{dF_{\rm coll}^{\rm II}}{dz} (z^{\prime}) \right]
\right\}
{\tilde V}_{\rm HII}(z^{\prime},z).
\label{eq:filling}
\end{eqnarray}
The second term on the right hand side
includes the factor $(1-F_{\rm HII})$ to allow for photo-ionization
feedback, while ${\tilde V}_{\rm HII}(z_{\rm on},z)$ is the volume per
unit mass and unit efficiency ionized by redshift $z$ by a single
source that turned on at an earlier redshift of $z_{\rm on}>z$. In
Eq.~(\ref{eq:filling}),
$\epsilon_{i} \equiv N_{\gamma} f_* f_{\rm esc}$
represents the overall efficiency of star-formation and photon
emission, where $f_*$ is the fraction of baryons in the halo that turns
into stars, $N_{\gamma}$ the mean number of ionizing photons produced
by an atom cycled through stars (averaged over initial mass
function), $f_{\rm esc}$ is the fraction of these photons escape to IGM.
In the following we use fiducial values of $\epsilon$ for type Ia and
Ib halos as advocated by HH. We set $\epsilon_{\rm II}=0$, i.e. we do not
consider star formation in ``minihalos'' in which the gas can cool 
via molecular hydrogen cooling. 
It is straightforward to include Type II halos in our models, but
the ionization history involving them is more complicated.
For our purpose -- test the effect of non-Gaussianity -- it is unnecessary
to introduce such additional complexity. Indeed, one aim 
is to see if early reionization could be obtained without
invoking molecular hydrogen cooling.

The evolution of the ionization
front is calculated by solving the following differential equation
\begin{equation}
\frac{dV}{dt} = 3H(z)V + \frac{\dot{N}_\gamma(t)}{\bar{n}_H} 
- C \bar{n}_{\rm H} \alpha_{\rm B} V \, ,
\end{equation}
where $\alpha_B=2.6 \times 10^{-13}$ cm$^3$ s$^{-1}$ is the case B
recombination coefficient. Adopting a fixed value of $\alpha_B$ is
equivalent to assuming a constant temperature of the ionized region.
Strictly speaking, this is not true, and one needs
consider the evolution of the thermal state of the ionized
IGM. However, with atomic cooling the gas in the ionized, photo-heated
regions quickly approaches this temperature, and as shown by HH this
provides a good approximation.
$C=\langle n_H^2\rangle /\bar{n}_H^2$ is
the clumping factor, which we take to be at a constant value of 10. 
$H(z)$ is the expansion rate, with
$H_0[\Omega_m(1+z)^3+\Omega_\Lambda]^{1/2}$, and $\dot{N}_\gamma(t)$
is the rate of ionizing photon emission as a function of time.  We
take $\dot{N}_\gamma(t)$ to be a constant for $t < 3.1 \times 10^6$
years with a value of 3.7 $\times$ 10$^{46}$ s$^{-1}$ per solar
mass of stellar content and decrease from this constant value as
$(t/3.1 \times 10^6 {\rm yr})^{-4.5}$ thereafter. We start integration at
a sufficiently high redshift, and solve
Eq.~(\ref{eq:filling}) up to $F_{\rm HII}=1$, at which point we consider 
reionization is completed and keep $F_{\rm HII}$ fixed at 1 for redshift
below this\footnote{Since this
set of equations does not include evolution of IGM thermal state, it
produce unphysical result ($F_{\rm HII}>1$) if integrated past complete
reionization, however the result is approximately correct if the
integration is limited to before the completion of reionization.}.
The optical
depth is calculated by integrating the ionized volume fraction weighted by the
number density such that
\begin{equation}
\tau(z)= 1.08 \int_0^z dz\; \frac{dt}{dz} c \sigma_T n_{\rm H}(z)F_{\rm HII}(z).
\end{equation}
The factor 1.08 accounts approximately
for the contribution of helium reionization (see HH).

\section{Non-Gaussianity}

The collapsed fraction can be calculated using a standard mass
function, such as the Press-Schechter theory (PS; Press \& Schechter
1974) based on Gaussian density fluctuations, or the numerical
simulated mass function by Jenkins et al. (2001), which was used by
HH. To describe the linear power spectrum, we use the transfer function from
Eisensetin \& Hu (1998).
We now modify the standard mass function by adding non-Gaussian fluctuations
to the density field. To compare
with various available constraints, we make use of a simple model for the
non-Gaussian mass function which has been previously studied in the
literature (Koyama, Soda \& Taruya 1999; Willick 2000), though we note that, under specific
models of primordial non-Gaussianity, more detailed mass functions can be constructed 
(e.g., Matarrese, Verde \& Jimenez 2000). We write the generalized form of the mass
function as
\begin{equation}
n(m,z) = \frac{\bar{\rho}}{m} \frac{\delta_c}{\sigma^2(m,z)}
\frac{d\sigma(m,z)}{dm} \psi \left(\frac{\delta_c}{\sigma(m,z)}\right)
\, ,
\end{equation}
where the probability distribution function for density fluctuation,
$\delta$, is written in terms of a dimensionless function $\psi(x)$:
\begin{equation}
P(\delta|\sigma) = \sigma^{-1} \psi
\left(\frac{\delta_c}{\sigma}\right) \, ,
\end{equation}
where $\int \psi(x) dx =1$, $\int x \psi(x) dx=0$ and $\int x^2
\psi(x) dx=1$. Here, $\sigma(M,z)$ is the rms mass function on a scale
corresponding to a mass $m$, and $\delta_c$ is the critical density for
collapse (see, e.g., Kitayama \& Suto 1996 for details).
The standard Press-Schechter formula assumes a Gaussian probability
distribution function (PDF) such that $\psi(x) = 2A e^{-x^2/2}$ and
$A=(2\pi)^{-1/2}$.  To include a mild non-Gaussian tail to the
Gaussian description, essentially a skewness, we consider a PDF which
is based on the Poisson description, but modified appropriately for
the continuous variables involved (see, Willick 2000 for details). The
function is written as
\begin{equation}
\psi(x) = 2 A \frac{\lambda^{\sqrt{\lambda}x+\lambda+0.5}e^{-\lambda}}{\Gamma(\sqrt{\lambda}x+\lambda+1)} \, ,
\end{equation}
such that $\lambda$, which is a function of redshift, is a free-variable 
that captures the non-Gaussian nature of the PDF; As $\lambda$ is increased, 
the mass function approaches to the PS value for Gaussian distribution of 
density fluctuations. $A$ is chosen to return the correct normalization: 
$\int m n(m,z) dm = \bar{\rho}(z)$.

Note that in our non-Gaussian mass function, skewness, $S_3$, scales as $\lambda^{-1}$, 
which explains why non-Gaussianity increases with decreasing $\lambda$. 
In terms of the moments of the primordial density field, one can write 
skewness as
\begin{equation}
S_3(m,z) = \epsilon \frac{\mu_3(m,z)}{\sigma^4(m,z)} \, ,
\label{eqn:skew}
\end{equation}
where $\mu_3$ is the third moment of the density perturbations, scaled to 
a redshift of $z$, when $\epsilon$ is the parameter related to the quadratic 
correction (Matarrese, Verde \& Jimenez 2000). If we ignore for now the mass-, 
or equivalently scale-dependence, of primordial non-Gaussianity, 
then $\mu_3 \sim 6 \sigma^4$, and assuming linear growth of fluctuations, 
$S_3(z) \sim 6 \epsilon/G(z)$ where $G(z)$ is the growth factor. The latter 
scaling can be understood based on the fact that $\mu_3(m,z)$ and $\sigma^2(m,z)$ 
scale with redshift as $G^3(z)$ and $G^2(z)$, respectively. Since in our mass 
function, $S_3 \sim \lambda^{-1}$, we expect $\lambda$ to behave as 
$\sim (\epsilon^{-1}/6) G(z)$ as a function of redshift.

\begin{figure*}
\centerline{\psfig{file=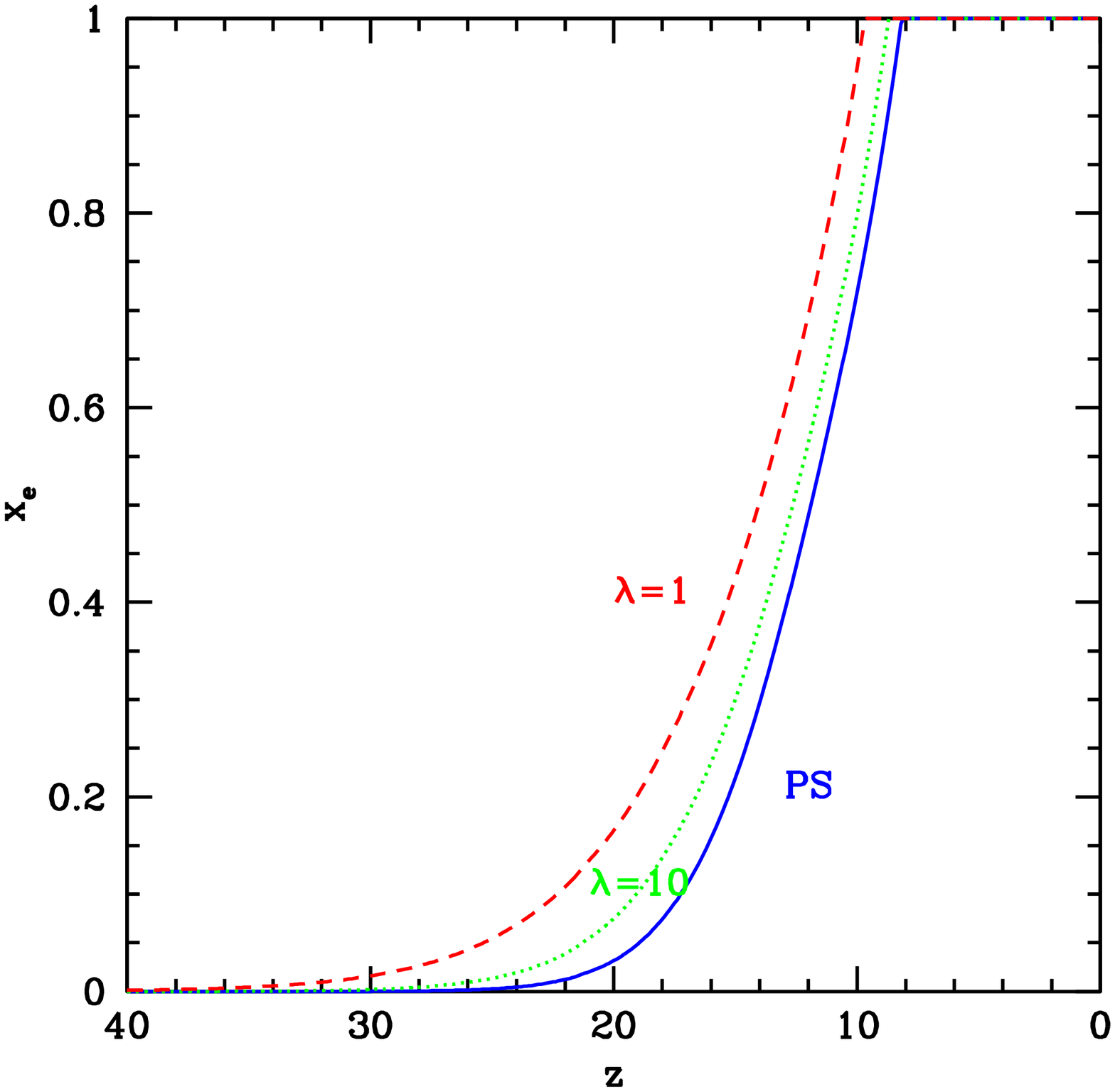,width=65mm,angle=0}
\psfig{file=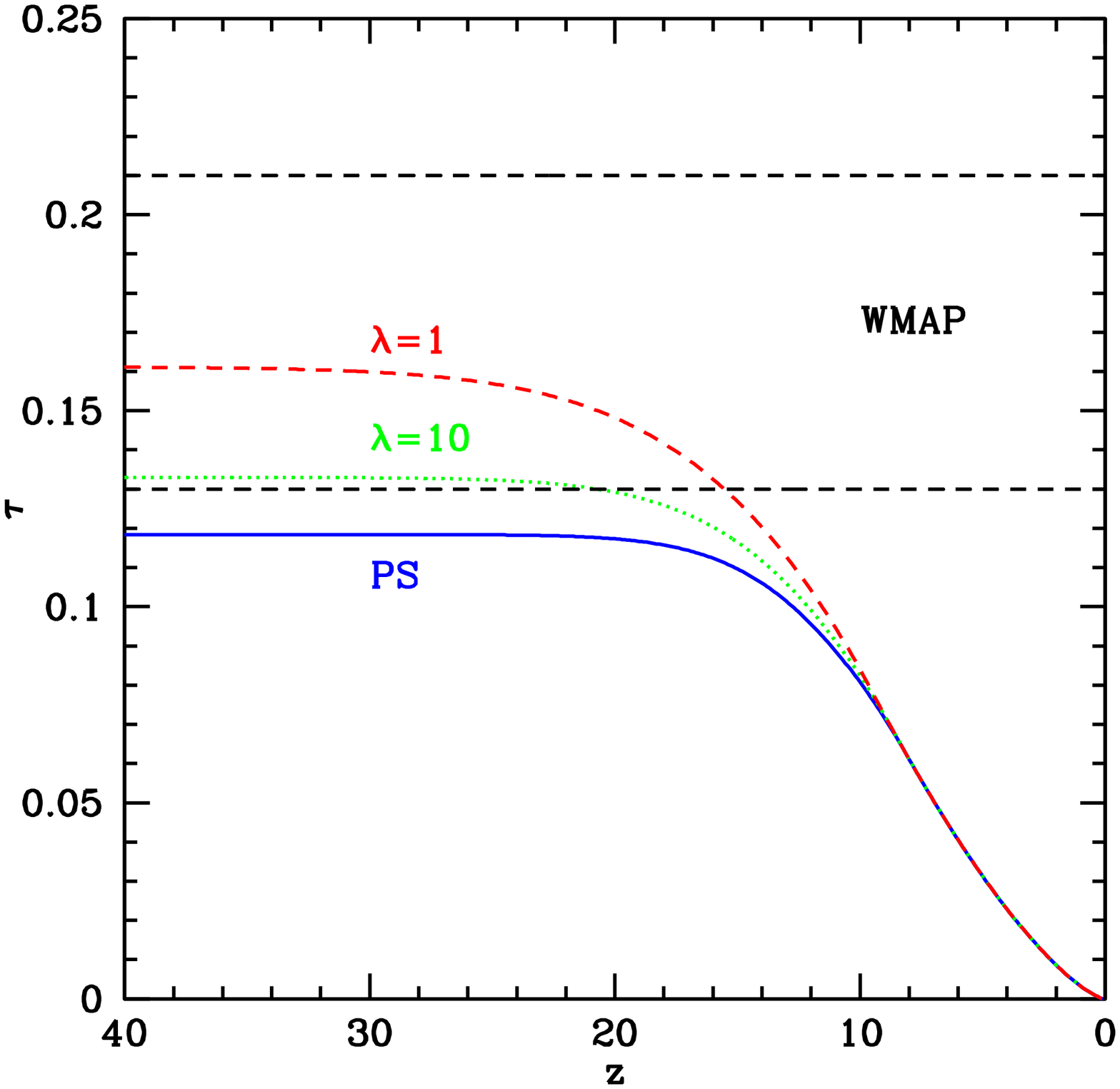,width=65mm,angle=0}
\psfig{file=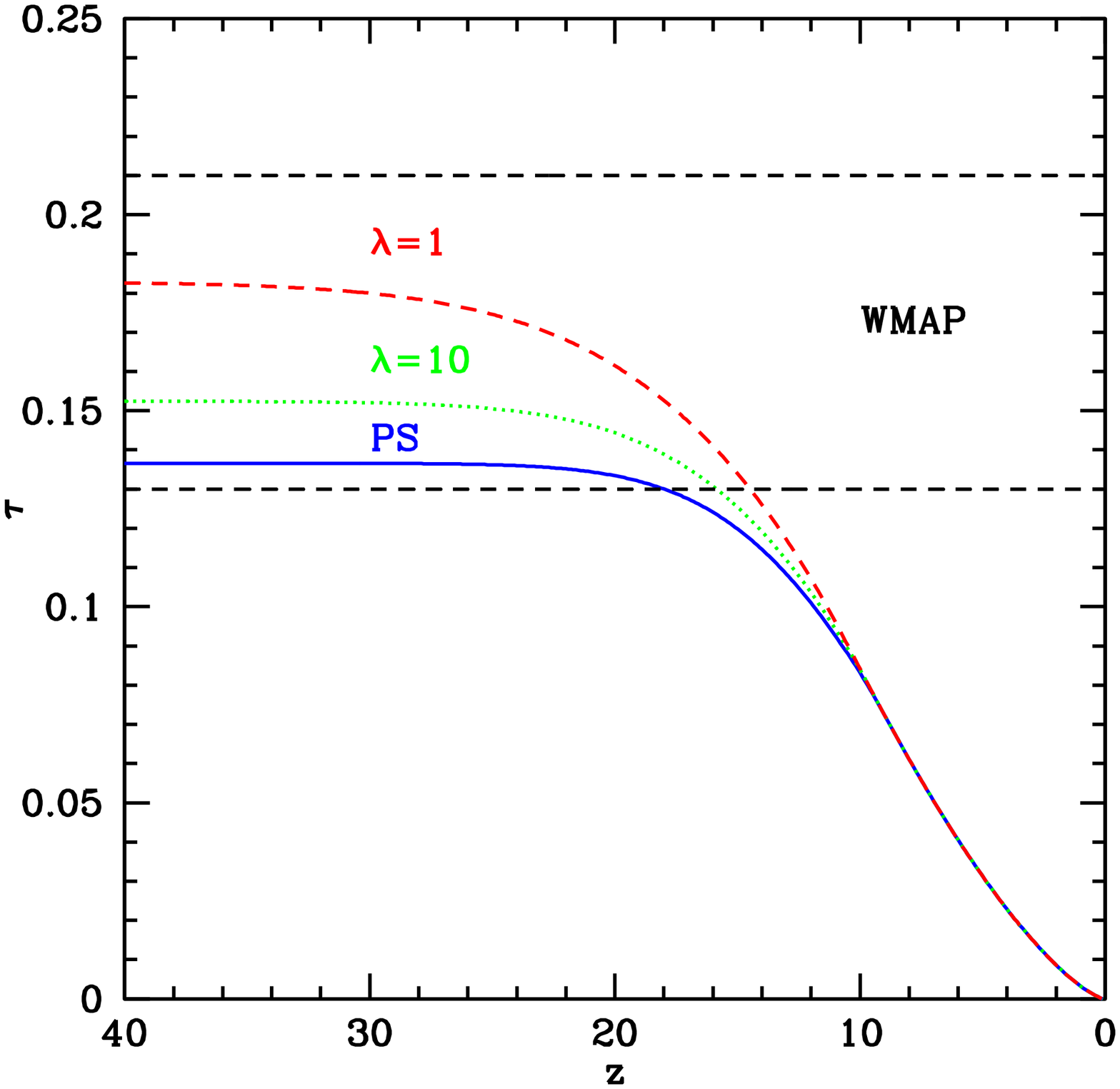,width=65mm,angle=0}}
\caption{{\it Left:} The volume ionized fraction,
$x_e(z) \equiv F_{\rm HII}(z)$ as a function of redshift. {\it Middle:}
The Thomson optical depth for electron scattering. The models are
based on ``ordinary'' star-formation in Type I halos only.
The Press-Schechter mass function leads to an optical depth with a value of
0.097 while the non-Gaussian mass functions lead to values of 0.11 and
0.13 for $\lambda=10$ and $\lambda=1$, respectively.
The two black dashed lines in the middle and right panels indicate the range of 
the first year WMAP result.
{\it Right:} Same as the middle plot, but with $\sigma_8=1$ and keeping 
all other parameters the same. }
\label{fig:tau}
\end{figure*}

In Fig.~1, we plot the collapsed fraction of these halos for the standard PS 
mass function and the non-Gaussian mass functions with $\lambda=10$ and $\lambda=1$
as a function of redshift.
We note that these values could be a function of redshift, although we have taken 
$\lambda$ to be a constant, for purposes of the present discussion. 
As $\lambda$ is decreased, or models become strongly non-Gaussian, the  
fraction of mass in collapsed halos increases at a fixed redshift. 
This illustrates the basic idea behind our calculation that with increasing non-Gaussianity of the
density field, there are more high overdensity peaks which collapse earlier. 
These early collapsing halos can form stars at an earlier time than in the
standard description and reionize the universe such that the total optical
depth is higher than previously estimated in the Gaussian models.

In Fig.~2, we show the ionization fraction and
optical depth for the models with the PS mass function and with the
non-Gaussian mass function with $\lambda=10$ and $\lambda=1$.
In addition to our standard cosmological values, to illustrate the
increase in the optical depth by halos collapsing at an earlier time,
we also calculate the optical depth in a model where $\sigma_8=1$
(Fig.2, right plot). We emphasize that such high values of $\sigma_8$ 
easily make the optical depth large even within the Gaussian model. 

For a given set of cosmological parameters,
higher optical depths can be achieved with the non-Gaussian mass
functions. As shown in Fig.~2, compared with the standard
Gaussian PS function, there is only a moderate increase in optical depth
for the $\lambda=10$ model (about 0.01 in $\tau$). This difference is in fact
comparable in size with the difference between the PS and Jenkins mass
functions: it is well known that compared with the PS mass function,
the Jenkins et al. (2001) function  predicts an excess of halos at the 
upper end of the halo mass.
Our calculated optical depth with the Jenkins et al. (2001) mass
function of $\sim$ 0.13 is consistent with previous calculations on
the optical depth based on Pop II stars alone (e.g., HH; 
Fukugita \& Kawasaki 2003). To increase the optical depth
beyond this value, one can either add more Population III stars at 
earlier epochs (Cen 2003; Wythie \& Loeb 2003; Sokasian et al. 2003) 
or, as we have considered, by considering smaller
$\lambda$ value (greater non-Gaussianity) while resorting 
only ordinary stellar populations.

A significantly non-Gaussian model with $\lambda \sim 1$ increases
the optical depth by a large fraction (about 0.04 in $\tau$). From Eq.~\ref{eqn:skew} 
and the expected scale with redshift, such a non-Gaussianity implies a 
primordial non-Gaussianity with $\epsilon \sim  5 \times 10^{-3}$.
Non-Gaussianities at this level can soon be tested with observations
of galaxy counts at redshifts of 5, though low redshift large scale 
structure observations will not be helpful for this purpose (Verde et al. 2001).
At present, according to Willick (2000) (see his Fig.~11), for 
$\Omega_m=0.27$, the low redshift galaxy cluster abundance
is consistent at the 90\% confidence level with a non-Gaussian mass
function of $\lambda < 14$. The cluster abundance at low-redshift is 
not a strong discriminator of non-Gaussian models, as argued by Verde et al. (2001). 
Note that only a constraint on the upper limit of $\lambda$ is available
from the cluster abundance in Willick (2000), though in Koyama et al. 
(1999; also, Robinson, Gawiser \& Silk 1998) $\lambda \sim 3$ is 
preferred for $\Lambda$CDM cosmological models. Again this is not a 
strong constraint of non-Gaussianity at redshifts of order 30.

Apart from low and moderate redshifts observations, non-Gaussianity 
can also be constrained, at a redshift of 1100, from observations
of cosmic microwave background anisotropies (e.g., Komatsu et al. 2003). 
Current constraints indicate a limit of 134 on $f_{\rm nl}$, at the 95\% 
confidence level; this parameter is related to the quadratic corrections to 
potential field and is different from $\epsilon$ we have been using to 
describe primordial non-Gaussianity based on corrections to the density 
field directly. Through the Poisson function, one can relate the two
parameters such that at large scales of order 100 Mpc, roughly, 
$f_{\rm nl} \sim 2 \times 10^3 \epsilon$, and the upper limit on $f_{\rm nl}$ of
134 from WMAP corresponds to $\epsilon < 7 \times 10^{-2}$ and allows 
for significant non-Gaussianity, rendering a lower limit on $\lambda$
at a redshift $\sim$ 30 of 0.08. While cluster counts from low redshifts 
have constrained the upper limit and CMB observations have constrained
the lower limit, the allowed range is still significant enough to 
produce substantial non-Gaussian fluctuations and to easily reproduce the
observed optical depth with early formation of structures and first stars.

Note that with completed WMAP results and with Planck, the limits on 
non-Gaussianity will be improved and, naively, small values for $\lambda$  at 
$z\sim 30$ may be ruled out. Such a limit, however, cannot fully
constrain non-Gaussianities, since one might still 
circumvent this by considering scale-dependent non-Gaussian models.
For example, one can apply the significant non-Gaussianity to scales which are related to
star-formation at redshifts above 10 and still allow such non-Gaussianity 
to remain today. It will then affect the number counts of $\sim 10^8$ M$_{\sun}$ 
halos at low redshifts, but CMB anisotropies at large scales will remain 
nearly Gaussian. The same models can also be tuned such that cluster counts 
at low redshifts remain Gaussian also. 
We note that it is not difficult to envision how such models can be constructed 
with examples including non-standard inflationary models
(Wang \& Kamionkowski 2000; Bartolo, Matarrese \& Riotto 2002; Bernardeau \& Uzan2003) 
and models with primordial black hole formation proposed by Bullock \& Primack 
(1997).
While constructing a concrete model for producing such scale-dependent 
non-Gaussian primordial fluctuations is beyond the scope of the present study,
we argue that these models can also be constrained eventually. The best way to 
probe the level of non-Gaussianity suggested above, and at scales related to 
this problem, may be to study number counts of galaxies at the highest redshifts 
possible (Verde et al. 2001).
Another possibility involves a measurement of the non-Gaussianity in the surface 
brightness fluctuations produced by first stars and proto-galaxies
in the near-infrared background and potentially detectable with upcoming deep 
observations from Near-IR Imaging Camera on the ASTRO-F mission (Cooray et al. 2003). 
Studying these constraints on the presence of non-Gaussianities at highest redshifts related to formation of
first objects is clearly needed to understand the extent to which the reionization 
process is affected.

\section{Summary}
We have examined how the observed high optical depth is achieved
in cosmological models with non-Gaussian primordial density fluctuations.
In these models, high over-density peaks due to the non-Gaussianity
collapse early on and may harbor first stars at redshifts above 
what is generally expected in the standard Gaussian model.
For mild non-Gaussian fluctuations at redshifts of 30, the increase in optical 
depth is at the level of 0.01, and it is unlikely to explain the observed value of
optical depth by this mechanism alone. To obtain significant increase in the optical depth,
the first collapsing halos associated with the primordial non-Gaussianity must contribute
to reionization significantly. A substantial volume fraction of the IGM
may be reionized if a number of Population III stars are formed in these early halos. 
We found that increasing the degree of
non-Gaussianities can easily reproduce the high optical depth.
Such models are still viable within constraints from 
both cluster number counts at low redshifts and CMB anisotropies at a redshift of 1100. 
While constraints from these two redshift ends will improve, models with scale-dependent 
non-Gaussianity are possible to explain the optical depth. Such models will only be constrained
by probes of non-Gaussianity related to mass scales of 10$^8$ M$_{\rm sun}$ at redshifts of 30 
and will involve, for example, study of first galaxy counts at very high redshifts.

 The ongoing operation of WMAP will pin down a presice value for the total optical depth.
For a longer-term, post-WMAP CMB polarazation experiments such as Planck will probe the reionization history
(Kaplinghat et al. 2003), and detection of the second-order polarization 
anisotropies can place additional constraints on details of the reionization process (Liu et al. 2001).
Data from these future observations will improve our picture of reionization 
and will enable us to further distinguish theoretical models related to the first generation of luminous structure formation.

\section*{Acknowledgments}

We thank organizers of the US-Japan Workshop on the SZ effect, where this paper was conceived, for a wonderful conference.
This work is supported by the Sherman Fairchild foundation and
DOE DE-FG 03-92-ER40701 (at Caltech) and by NSF PHY99-07949 (at KITP).
NY acknowledges support from Japanese Society of Promotion of Science
Special Research Fellowship (Grant 2674).
NS is supported by Japanese  Grant-in-Aid for Science Research Fund of the
Ministry of Education, No.14340290.

\end{document}